\documentclass[aps,prb,twocolumn,nopacs,superscriptaddress,longbibliography]{revtex4-1}
\usepackage[breaklinks=true,colorlinks,citecolor=blue,linkcolor=blue,urlcolor=blue]{hyperref}
\usepackage{graphicx,dcolumn,bm,multirow,hyperref,mathtools,braket,color,url,epstopdf,amssymb,amsmath}
\def\A100{[10\={1}0] }
\def\B120{[1\={2}10]}

\begin{document}

\title{Weak antilocalization and Shubnikov$-$de Haas oscillations in CaCuSb single crystal}

\author{Souvik Sasmal}
\email{sasmalsouvik6@gmail.com}
\affiliation{Department of Condensed Matter Physics and Materials Science, Tata Institute of Fundamental Research, Homi Bhabha Road, Colaba, Mumbai 400 005, India.}

\author{Vikas Saini}
\affiliation{Department of Condensed Matter Physics and Materials Science, Tata Institute of Fundamental Research, Homi Bhabha Road, Colaba, Mumbai 400 005, India.}

\author{Nicolas Bruyant}
\affiliation{Laboratoire National des Champs Magnétiques Intenses, CNRS-UGA-UPS-INSA, 31400 Toulouse, France}

\author{Rajib Mondal}
\affiliation{UGC-DAE Consortium for Scientific Research, Kolkata Centre, Bidhannagar, Kolkata 700 106, India.}

\author{Ruta Kulkarni}
\affiliation{Department of Condensed Matter Physics and Materials Science, Tata Institute of Fundamental Research, Homi Bhabha Road, Colaba, Mumbai 400 005, India.}

\author{Bahadur Singh}
\email{bahadur.singh@tifr.res.in}
\affiliation{Department of Condensed Matter Physics and Materials Science, Tata Institute of Fundamental Research, Homi Bhabha Road, Colaba, Mumbai 400 005, India.}

\author{Vikram Tripathi}
\affiliation{Department of Theoretical Physics, Tata Institute of Fundamental Research, Homi Bhabha Road, Colaba, Mumbai 400 005, India.}

\author{A. Thamizhavel}
\email{thamizh@tifr.res.in}
\affiliation{Department of Condensed Matter Physics and Materials Science, Tata Institute of Fundamental Research, Homi Bhabha Road, Colaba, Mumbai 400 005, India.}


\begin{abstract}

Quantum oscillations in both linear and Hall resistivities and weak antilocalization (WAL) are barely observed in bulk single crystals. Here we report the transport properties of a CaCuSb single crystal that crystallizes in the hexagonal crystal structure. The magnetotransport studies reveal WAL and Shubnikov$-$de Haas (SdH) quantum oscillations with a unique frequency at $314$~T. A cusp-like behavior in the low field regime of magnetotransport for $J~\parallel~(ab)$-plane and $B~\parallel$~[0001] confirms the WAL in CaCuSb. Angular-dependent normalized magnetoconductance and SdH oscillations studies reveal that the observed phenomena originate from the $2D$ transport channels. The high magnetic field (up to $45$~T) experiments demonstrate plateau-like features in the Hall measurements. The first-principles calculations unfold that CaCuSb is a non-topological semimetal with dominant hole carries at the Fermi level. Our study reveals that CaCuSb is a promising candidate to explore the quasi$-2D$ quantum transport phenomenon in the transition metal pnictide materials. 

\end{abstract}

\maketitle

%
\section{INTRODUCTION}
 Since the discovery of topological insulators (TIs) with nontrivial metallic surface states~\cite{RevModPhys.82.3045, Moore2010}, substantial research efforts are devoted in searching new topological semimetals owing to their novel electronic structure and interesting physical properties~\cite{doi:10.1146/annurev-matsci-070218-010023,Burkov2016,https://doi.org/10.1002/adma.201704382,https://doi.org/10.1002/adma.201702359,Berger2018,Wu2019,https://doi.org/10.1002/adfm.201803746,Cai2018}. Following the foot-print of graphene~\cite{Meyer2007}, $2D$ materials are widely investigated due to their exotic $2D$ transport properties which seed from carrier confinement effects~\cite{https://doi.org/10.1002/adma.201901694,https://doi.org/10.1002/adfm.201904932}. It has been found that in addition to $2D$ materials, the topological materials with layered crystal structure also show the $2D$ behavior of carriers. In this regard, the transition metal-pnictides layered ternary compounds of $ABC$-type offer a prominent setup for realizing exotic topological and transport properties\cite{Singh_B2018,Singh_MR2019, Sasmal_2020}. For example, CaAgBi~\cite{Sasmal_2020} exhibits Dirac-Weyl semimetal with weak antilocalization (WAL) effects, whereas LuPdBi~\cite{Pavlosiuk2015} and YPtBi~\cite{PhysRevB.84.220504} exhibit WAL as well as superconductivity at low temperature.  When the phase coherence length ($l_{\Phi}$) is larger than the mean free path ($l$), quantum interference results in either the weak localization (WL) or WAL effects depending on whether the interference is constructive or destructive, respectively~\cite{10.1117/12.2063426, datta_1995, Zhao2019}. The strong spin-orbit coupling (SOC) effects lead to spin-rotation in materials whenever an electron gets scattered due to impurity, and thereby causing the phase change of the electron wave function, realizing a destructive interference. The destructive interference enhances the conductivity and a cusp-like negative conductivity is observed in the single crystals of various ternary materials~\cite{PhysRevB.95.195113, Sasmal_2020, PhysRevB.99.241102}. At a sufficiently high magnetic field, if the system is in a quantum diffusive regime ($\hbar\omega>k_BT$), the Fermi energy crosses the Landau levels with increasing field, leading to Shubnikov$-$de Haas (SdH) oscillations in the magnetoresistance. There are a few reports that show both the WAL and quantum oscillations in longitudinal resistivity, although oscillations in Hall data are usually absent in those materials ~\cite{Bao2012, Bhardwaj2018}.
 
 In this study, we report magnetotransport properties and demonstrate the WAL and SdH quantum oscillations in both the linear and Hall resistivities in a single crystal of ${\rm CaCuSb}$. CaCuSb belongs to a large family of $ABC$-type hexagonal materials with the space group $P6_3/mmc$ (No. 194). A cusp-like behavior is observed in the low field region of magnetoresistance (MR), confirming the WAL effect in CaCuSb. At sufficiently high magnetic fields, SdH oscillations are observed which enable us to reveal its conduction and Fermi surface properties. Interestingly, the Hall effect measurements show the signatures of Hall plateaus at definite intervals. Our analyses suggest that the observed transport phenomena in $\rm{CaCuSb}$ are due to the quasi-$2D$ transport channels in the layered crystal structure, consisting of Cu-Sb layers that are separated by Ca layers.  


\section{Experimental Methods}

Good quality single crystals of CaCuSb were synthesized by flux method using Cu:Sb eutectic composition as a flux. High purity starting elements of Ca (99.98\% ingot from Alfa Aesar), Cu (99.999\% shot from Alfa Aesar) and Sb (99.999\% shot from Alfa Aesar) were taken in 1:4.5:7.5 molar ratio in an alumina crucible. The alumina crucible was sealed in a quartz ampoule under a partial pressure of Ar gas. The sealed ampoule was placed in a box type furnace, and heated  to 1050$^{\circ}$C at the rate of 30$^{\circ}$C/h and held at this temperature  for 24 h, for proper homogenization. Then the furnace was allowed to slow cool at the rate of 2$^{\circ}$C/h to 630$^{\circ}$C at which point the flux was removed by  centrifuging.  Electrical transport studies have been performed on a cut and aligned bar-shaped sample with  current along the principle crystallographic directions $viz.,$ in the basal plane and along the $c$-axis. Electrical transport measurements were carried out on a sample with dimension $1.5\times1\times0.12$~mm$^3$, using Quantum Design Physical Property Measurement System (PPMS) instrument with magnetic field up to 14 T, in the temperature range $2 - 300$~K.  High magnetic field Hall effect measurement was performed at the European Magnetic Field Laboratory (EMFL) at the Laboratoire National des Champs Magnétiques Intenses (LNCMI) at Toulouse, France, up to a mangetic field of $45$~T. 

\section{Results and Discussions}
\subsection{X-ray diffraction}

\begin{figure}[!]
\includegraphics[width=0.5\textwidth]{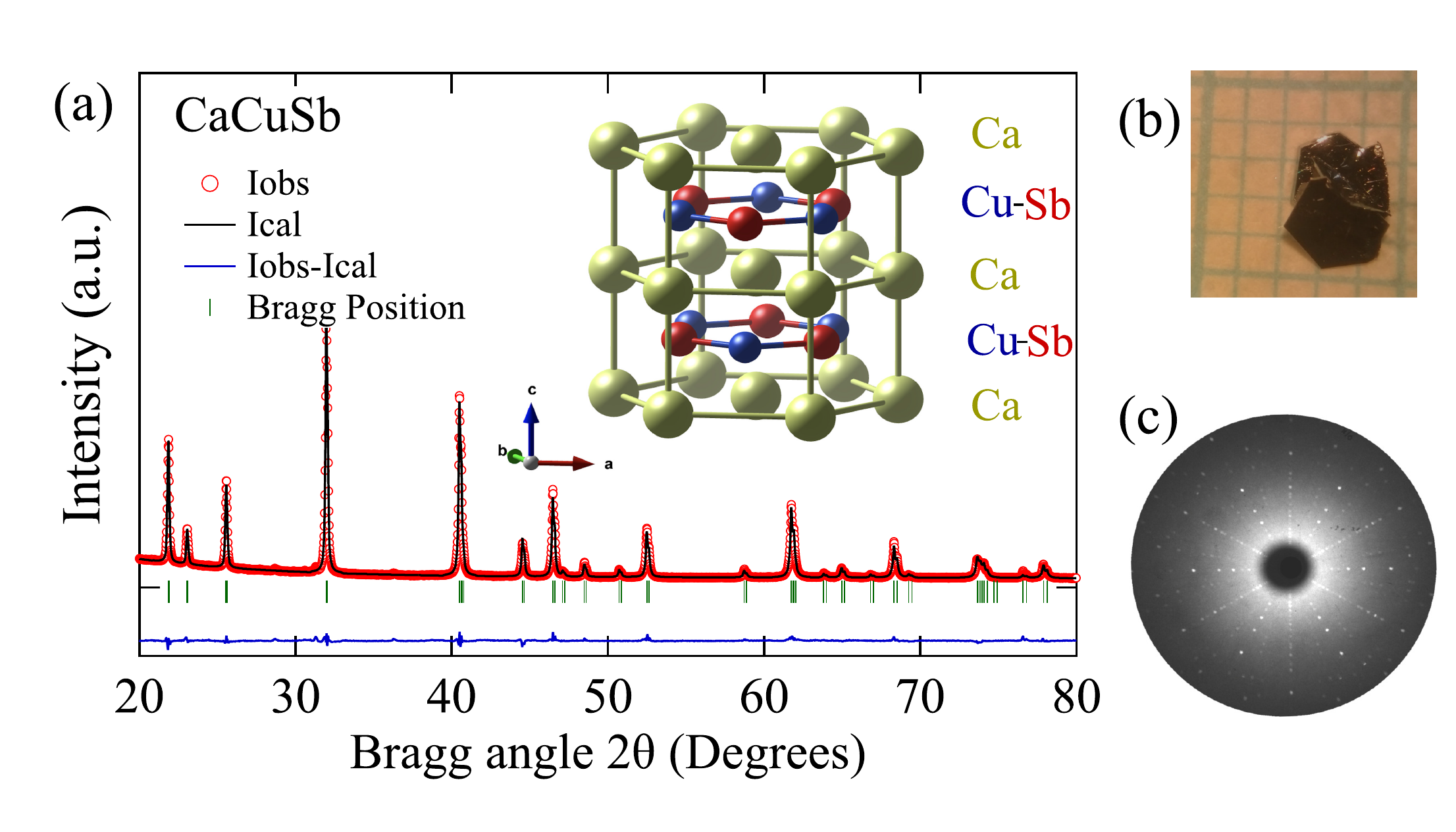}
\caption{(a) Powder x-ray diffraction pattern along with the Rietveld refinement of CaCuSb. Inset:  The hexagonal crystal structure of CaCuSb.  Each Cu$-$Sb layer is rotated by 60$^{\rm \circ}$ relative to the neighbouring Cu$-$Sb layer. (b)  As grown crystal image (c) Laue diffraction pattern of CaCuSb depicting the (0001)-plane.}

\label{Fig1}
\end{figure}

The room temperature powder x-ray diffraction (XRD) and Rietveld refinement of the XRD data are shown in Fig.~\ref{Fig1}(a).  From the refinement we estimate the lattice parameters as $a~=~4.452$~\AA~ and $c~=~8.133$~\AA~ which are in good agreement with the previously published data~\cite{Eisenmann}.  The Rietveld analysis also confirmed the space group as $P6_3/mmc$ (No. 194) and the Wyckoff positions of Ca, Cu, and Sb as $2a$, $2d$ and $2c$ respectively.  The crystal structure of CaCuSb is shown in the inset of Fig.~\ref{Fig1}(a).  It is evident that the crystal structure is layered along the hexagonal $c$-axis where Cu$-$Sb layers are sandwiched between Ca-layers and each Cu$-$Sb layer is rotated by $60^{\circ}$ relative to the neighbouring layer.  These Cu$-$Sb layers may act as a unique charge transport layers which are confined in the $(ab)$-plane.  As grown single crystal with the flat plane corresponding to (0001) is shown in Fig~\ref{Fig1}(b).  The Laue diffraction pattern of the flat plane is shown in Fig.~\ref{Fig1}(c).  Well defined Laue diffraction pattern together with the $6$-fold symmetry confirms the good quality of the single crystal.  

\begin{figure*}[!]
	\includegraphics[width=0.9\textwidth]{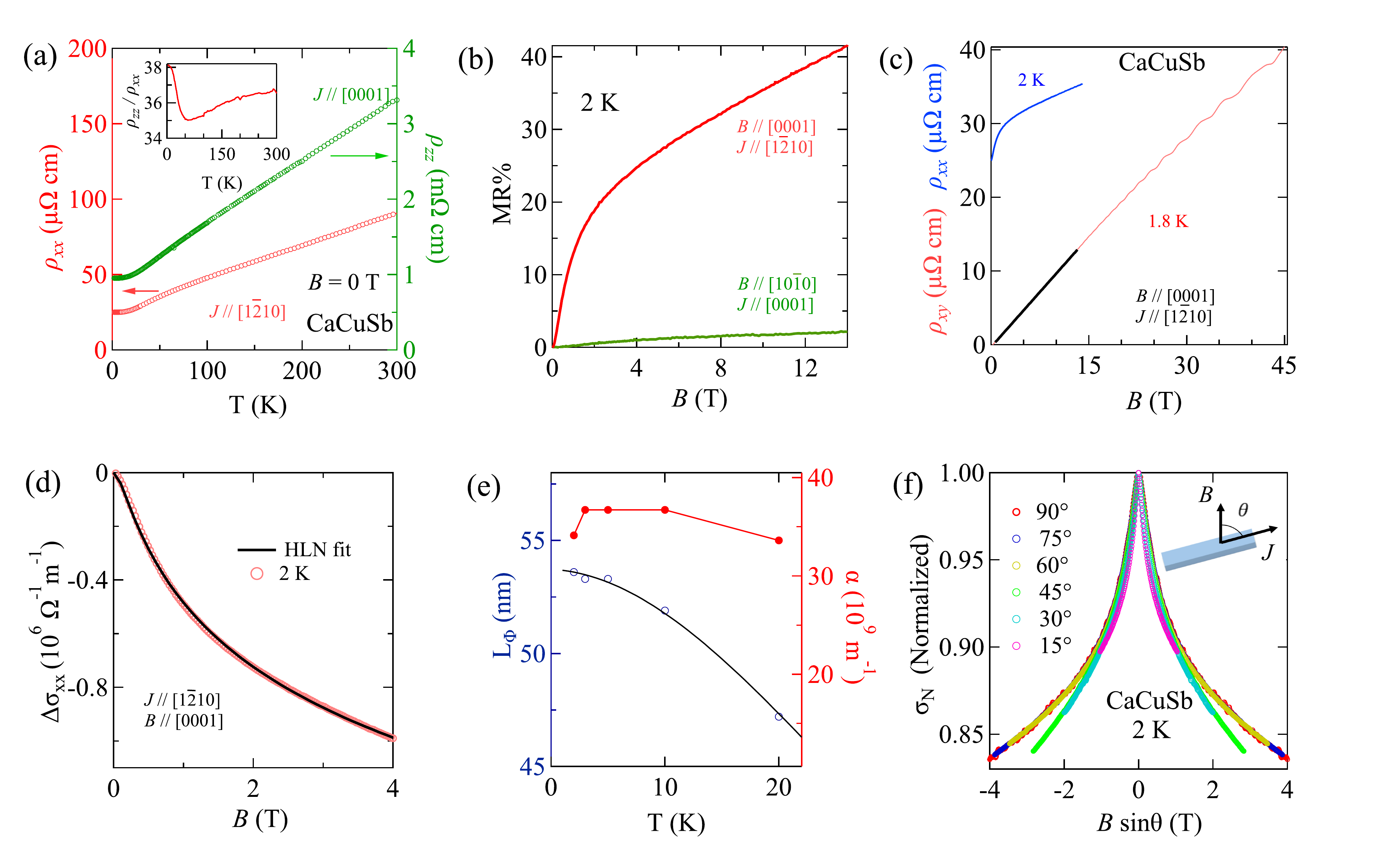}
	\caption{(a) Temperature dependence of the electrical resistivities $\rho_{xx}$ and $\rho_{zz}$ of CaCuSb in zero magnetic field.    The inset shows the anisotropy ratio ($\rho_{zz}$/$\rho_{xx}$) as function of temperature. (b) The magnetoresistance  at $T = 2$~K, for $J~\parallel~$[1\={2}10] and $J~\parallel~[0001]$. A cusp like behavior in low field region indicates the WAL effect. (c) Hall resistivity $\rho_{xy}$ as a function of magnetic field ($B$) at $1.8$~K. The black solid line show the linear fit to the data to extract the carrier concentration. The linear resistivity $\rho_{xx}$ up to $14$~T is also shown. (d) Magnetoconductivity ($\Delta\sigma_{xx}$) at $2$~K (red markers) and HLN fit (dash black line) (see text for details). (e) Phase coherence length ($L_{\Phi}$) and coefficient $\alpha$ as a function of temperature. Black solid line indicates the phase coherence length fitting using Eq.(~\ref{Eqn2}). (f) Normalized magnetoconductivity ($\sigma_{\rm N}$(Normalized)) at different tilt angles ($\theta$) between current ($J$~$\parallel$~[1\={2}10]) and $B$.}
	\label{Fig2}
\end{figure*}

\subsection{Electrical transport and Hall resistivity}

The temperature dependence of electrical resistivity is shown in Fig~\ref{Fig2}(a) for both in-plane ($\rho_{xx}$) and out-of-plane ($\rho_{zz}$) current directions which shows a metallic nature. The electrical resistivity is anisotropic depending on the direction of the current flow and a large anisotropy is observed for in-plane and out-of-plane current directions, thus it refers to a quasi-$2D$ nature of the electrical transport. Figure~\ref{Fig2}(b) shows the magnetoresistance (MR) at $T = 2$~K, for $J~\parallel~$[1\={2}10] ($ab$-plane) and $B~\parallel~$[0001], showing a cusp-like behavior that increases rapidly at low fields and linearly above $3$~T, and the MR reaches 41\% at 14~T. This cusp-like behavior in MR is reminiscent of the WAL effect seen in similar equiatomic systems~\cite{PhysRevLett.106.166805, PhysRevLett.108.036805,Sasmal_2020,Pavlosiuk2015,PhysRevB.84.220504}. Generally, the WAL phenomenon is observed due to $2D$ charge diffusion in the system. However, the presence of such WAL for $J~\parallel$~[1\={2}10] here suggests the possibility of destructive interference in the $ab$- plane. For $J~\parallel$~[0001] and $B~\parallel$~[10\={1}0], MR is 3.5\% at 14~T (12 times smaller than $ab$-plane) and almost linear in the entire field range without any sign of WAL in this direction, indicating the absence of such interference effects. These results support that the observed WAL has a $2D$ nature. The WAL behavior vanishes for $T > 50~K$, as shown in the Supplemental Material Figure S1~\cite{supplement}. Furthermore, in the range of $10-14$~T, SdH oscillation has been observed for $J~\parallel$~[1\={2}10] and $B~\parallel$~[0001], but it is absent for $J~\parallel$~[0001] and $B~\parallel$~[10\={1}0]. Here, we have symmetrized the linear resistivity $\rho_{xx} = [\rho_{xx}^{raw}(+B) + \rho_{xx}^{raw}(-B)]/2$ and antisymmetrized the Hall resistivity $\rho_{xy} = [\rho_{xy}^{raw}(+B) - \rho_{xy}^{raw}(-B)]/2$. The Hall resistivity is measured in a pulsed magnetic field of $45$~T at $T = 1.8$~K is shown in Fig.~\ref{Fig2}(c), which remains positive, suggesting that the hole carriers are dominant in the system. From the slope of Hall resistivity we estimate the carrier density and the mobility using a single band model as $n_{\rm h} \sim 5.81~\times~10^{26}$~m$^{-3}$ and $\mu_{\rm h} \sim 0.043$~m$^2$V$^{-1}$s$^{-1}$, respectively. The two-band model gives contribution from both the electron and hole (See Supplemental Material Figure S2 ~\cite{supplement}). 

\subsection{Weak antilocalization}

The calculated magnetoconductivity (MC), $\Delta \sigma_{xx}(B) = \sigma_{xx}(B)-\sigma_{xx}(0)$ with $\sigma_{xx}(B) = \frac{\rho_{xx}(B)}{\rho_{xx}^2(B)+\rho_{xy}^2(B)}$, for $J~\parallel~$[1\={2}10] and $B~\parallel~[0001]$ is shown in Fig.~\ref{Fig2}(d). It is obvious that the calculated MC also exhibits a cusp-like feature suggesting WAL in this quasi-$2D$ CaCuSb. Usually, this cusp-like feature can be seen at very low field for low dimensional systems. In the present case~at $B = 4$~T, the value of  $\omega\tau=\frac{eB}{m^*}\tau=0.26$, which is less than 1,  where $m^*=0.4 m_e$ and $\tau=1.5\times10^{-13}$~s, and $e$ is the electronic charge  (the quantum parameters $m^*$ and $\tau$ have been obtained from the SdH oscillations, as discussed later). This suggests the charge transport is in quantum diffusive regime,  so that WAL has been observed even at $B = 3$~T. Notably, the observed WAL in MC at low fields can be explained based on the Hikami-Larkin-Nagaoka (HLN) model~\cite{10.1143/PTP.63.707}. In the strong SOC limit, the HLN model is given as 
\begin{equation} 
\label{Eqn1}
\Delta \sigma_{xx} (B) = - \frac{\alpha e^2}{2 \pi^2 \hbar} \left[ \psi \left(\frac{\hbar}{4 e L_\phi^2 B} + \frac{1}{2} \right) \\ -ln \left(\frac{\hbar}{4eL_\phi^2B} \right)  \right],
\end{equation}
where $e$, $\psi$, and $L_{\phi}$ are the electronic charge, digamma function, and phase coherence length, respectively. From this HLN fit, the phase coherence length is estimated as $L_{\phi} = 53.6$~nm and ${\alpha} = 3.41~\times~10^{10}$~m$^{-1}$ (shown in in Fig~\ref{Fig2}(d)).  Here, $\alpha$ indicates the number of conduction channels per unit length. Considering the sample thickness of 120 $\mu m$, the obtained total conduction channels in the system is $\sim 4\times10^6$.  Such a large number of conduction channels  has been reported in several semimetallic systems like  Bi$_2$Te$_3$~\cite{PhysRevB.95.195113}, YbCdGe~\cite{PhysRevB.99.241102}, LuPdBi~\cite{Pavlosiuk2015}, YPtBi~\cite{pavlosiuk2016superconductivity}. Temperature dependence of phase coherence length ($L_{\Phi}$) is shown in Fig~\ref{Fig2}(e). Because of phononic contribution the $L_{\Phi}$ decreases with increasing temperature and it can be described by the relation
\begin{equation}
\label{Eqn2}
\frac{1}{L_{\Phi}^2(T)}=\frac{1}{L_{\Phi}^2(0)}+A_{ee}T+A_{ep}T^{2},
\end{equation}

 where $L_{\Phi} (0)$ the zero temperature phase coherence length, and $A_{ee}T$ and $A_{ep}T^{2}$  are the electron-electron and electron-phonon contribution, respectively. The obtained fitting parameters are $L_{\Phi}(0)=53.7$~nm, $A_{ee}=3.35\times10^{-7}$~nm$^{-2}K^{-1}$ and $A_{ep}=2.32\times10^{-7}$~nm$^{-2}K^{-2}$. The coefficient $\alpha$ is almost constant through the temperature regime $2-20$~K, this suggests the number of conduction channels are almost unaltered with temperature. Furthermore, we have also performed the angular dependence of MC at $T = 2$~K to understand the origin of WAL (Fig.~\ref{Fig2}(f)). The normalized MC $\sigma_{N}$(Normalized)$=\frac{R(0)}{R(B,\theta)}$ is plotted as a function $B\cdot{\rm sin}(\theta)$ at different tilt angles $\theta$ ($\theta$ is the angle between the direction of current and magnetic field).  The $\sigma_{\rm N}$ vs $B \cdot {\rm sin}{(\theta)}$ plot shows that the low field MC falls onto a single curve for all the measured field angles.  This suggests that WAL effect is originated from $2D$ parallel conduction channels in the bulk \cite{PhysRevB.83.241304,PhysRevB.95.195113,Xu2014}.  Similar behavior of MC curve is seen in other measured samples (Supplemental Material Figure S3~\cite{supplement}).  The $2D$ parallel conduction channels are presumably due to the Cu$-$Sb layers which is further confirmed from the band structure studies (to be discussed later) where the density-of-states (DOS) of the Cu and Sb atoms is much higher than Ca atoms at the Fermi level. 
 
\subsection{Shubnikov$-$de Haas (SdH) quantum oscillations}

\begin{figure}[!]
	\includegraphics[width=0.5\textwidth]{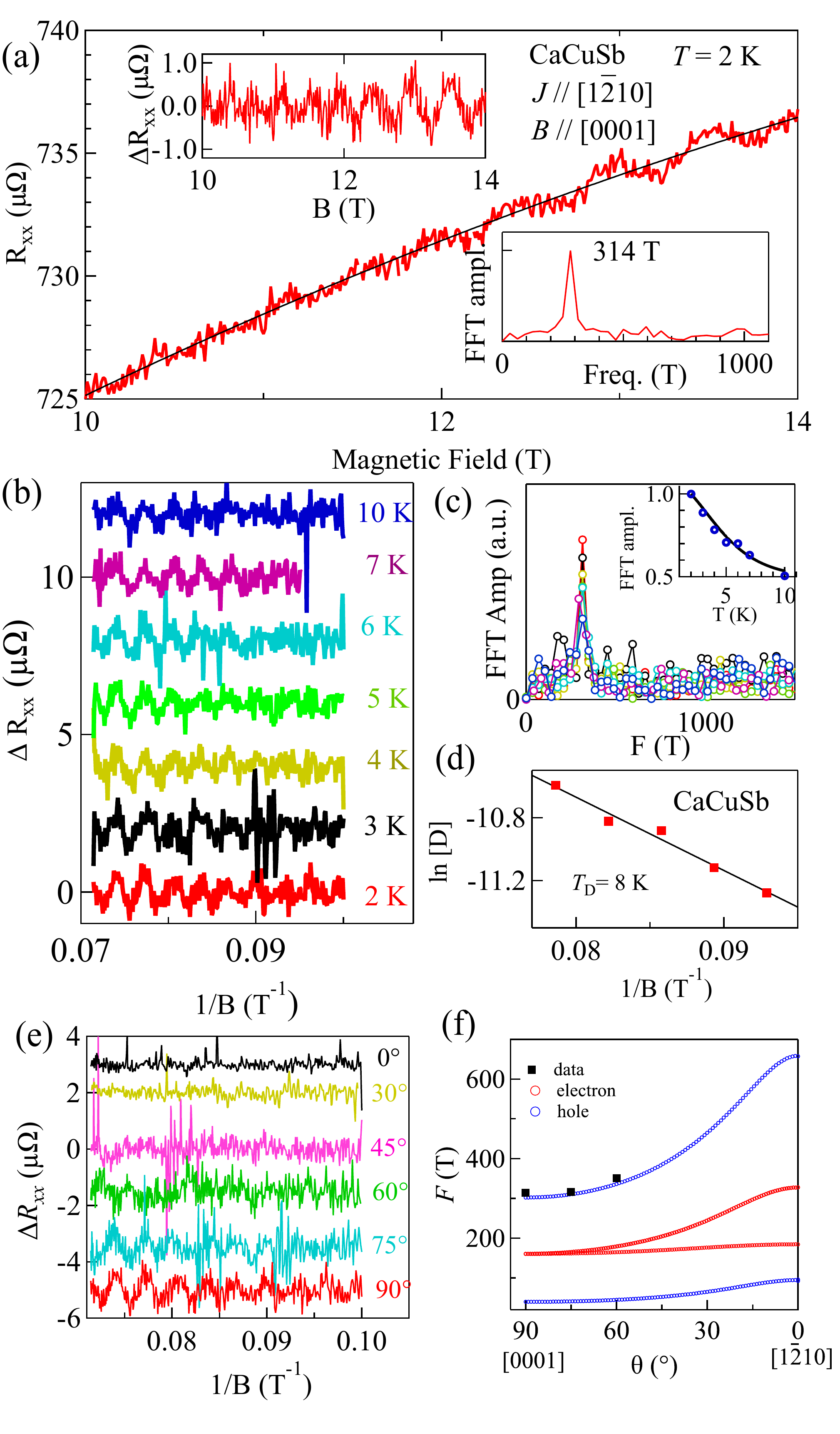}
	\caption{(a) SdH quantum oscillation at $2$~K. The solid black line shows the estimated background using a polynomial function. The top inset shows the background-subtracted SdH oscillations and the lower inset illustrates observed single quantum oscillation frequency at $314$~T. (b) SdH oscillation at different temperature (c) oscillation frequency at different temperature. The corresponding mass plot is shown in the inset. (d) Dingle plot ($T_{\rm D} = 8 $~K) for estimating the Dingle temperature in CaCuSb. (e)  $\Delta R_{xx}$ versus $1/B$ at various field angles. (f) Angular-dependent SdH oscillation frequencies of experimental data (black solid square), calculated electron (red) and hole (blue) pockets.   }
	\label{Fig4}
\end{figure}

It is interesting to note that the longitudinal resistance of CaCuSb at sufficiently large magnetic fields ($B~\parallel~[0001]$) oscillates periodically in $1/B$, depicting SdH quantum oscillations. The raw SdH oscillation observed in the field range $10-14$~T at $T=2$~K is shown in Fig.~\ref{Fig4}(a). We further obtain SdH frequency by subtracting the background signal using a polynomial function and plot $\Delta R_{xx}$ vs $B$ in the top inset of Fig.~\ref{Fig4}(a). A simple fast Fourier transform (FFT) of $\Delta R_{xx}$ as a function of $1/B$ reveals a unique frequency at $314$~T. A similar quantum oscillation in linear resistivity has been observed on a different sample B2, see Figure S4 in Supplemental Material~\cite{supplement}. From the Onsager relation $F = (\Phi_0 / 2 \pi^2) A_{ext}$ we have estimated extremal area $A_{ext}$ of the Fermi surface, where $\Phi_0$ is quantum of flux. $\Delta R_{xx}$ and FFT amplitude at different temperatures are shown in Fig~\ref{Fig4}(b)-(c). The oscillating part of the $\Delta R_{xx}$ has been analyzed using the standard Lifshitz-Kosevich (LK) equation which is given by~\cite{shoenberg2009magnetic}



\begin{equation}
\label{eqn3}
\Delta R_{xx} (T,B)\propto ~ R_{\rm B}~R_{T}~{\rm cos}\left[2 \pi\left(\frac{F}{B}+\frac{1}{2}+\beta \right)\right].
\end{equation}

\begin{figure}[!]
	\includegraphics[width=0.5\textwidth]{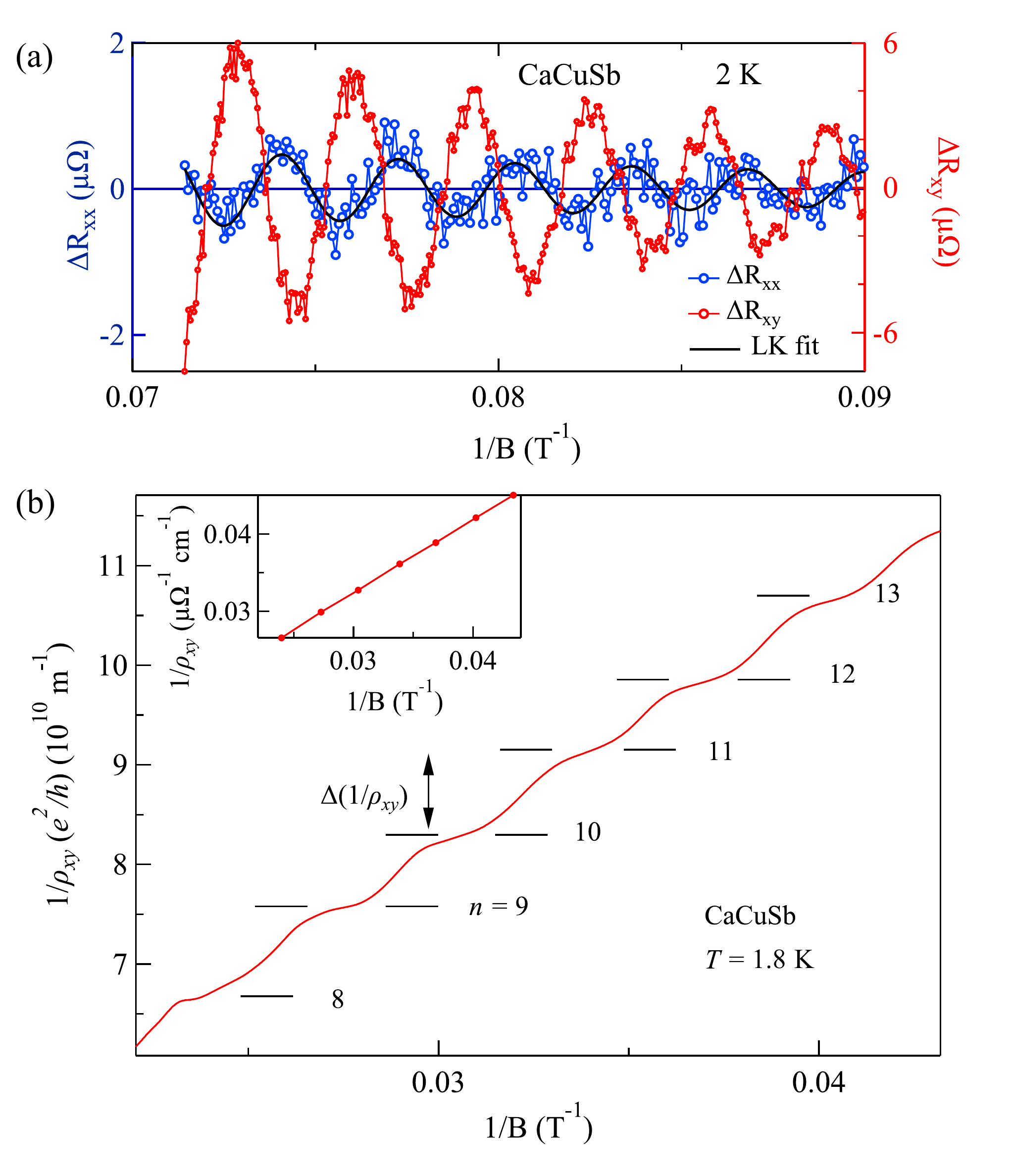}
	\caption{(a) $\Delta R_{xx}$ and $\Delta R_{xy}$ as a function of $1/B$. $\Delta R_{xx}$ is 90$^{\circ}$ out-of-phase with $\Delta R_{xy}$.  (b) $1/\rho_{xy}$ vs $1/B$ plot (measured up to a field of 45 T). The horizontal lines mark the Hall plateaus with LL index $n$.  Inset shows the plateau position of $1/\rho_{xy}$ as a function of $1/B$. The linear behavior  establishes the quantization nature of Hall data.  }
	\label{Fig5}
\end{figure}

\begin{table*}[ht]
	\centering  
	\caption{List of parameters estimated from the SdH quantum oscillations measurements. See the main text for the details.}
	\begin{ruledtabular}
		\begin{tabular}{c c c c c c c }
			$F ({\rm T})$    &   $A_{ext} ({\rm m}^{-2}) $     &      $\mu_{SdH} ({\rm m}^{2} {\rm V}^{-1} {\rm s}^ {-1})$  & $m^* (m_e)$   & $T_D ( {\rm K})$ & $\tau~({\rm s})$   \\
			\hline \\
			$314$    &   $3 \times 10^{18}$  & 0.0652                 & 0.4    & 8  & 1.5$\times 10^{-13}$
		\end{tabular}
	\end{ruledtabular}
	\label{Table1} 
\end{table*}

\begin{figure}[!]
\includegraphics[width=0.5\textwidth]{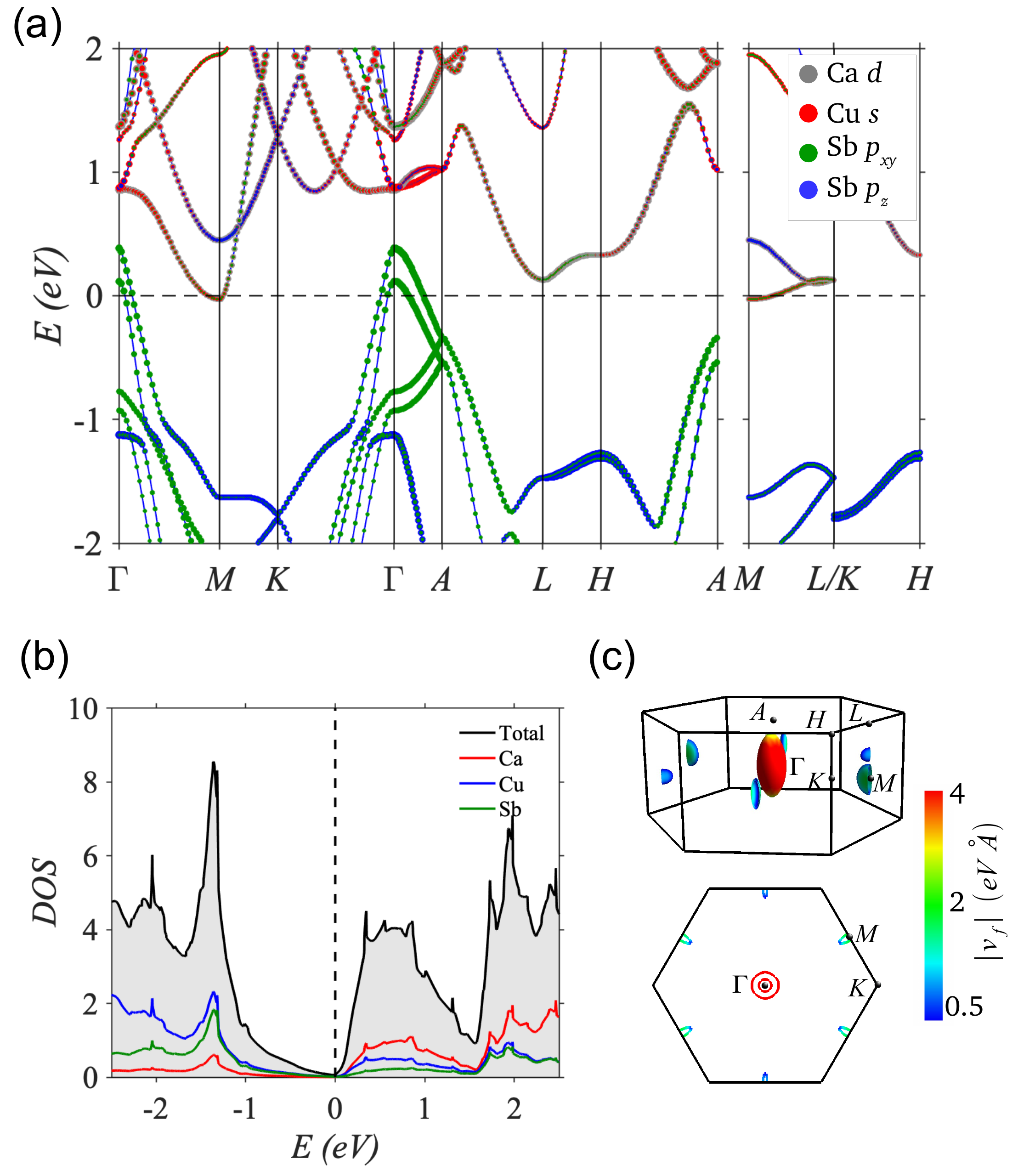}
\caption{(a) First-principles bulk band structure of CaCuSb with SOC. Sizes of gray, red, green, and blue markers represent contributions from Ca $d$, Cu $s$, Sb $p_{xy}$, and Sb $p_z$ states to various bands. (b) The total and partial DOS calculated with SOC. Black, red, blue, and green lines mark total, Ca, Cu, and Sb DOS, respectively. (c) Fermi surface with the Fermi velocity of associated electronic states. The blue and red regions highlight Fermi surface area with  low and high Fermi velocity, respectively.}
\label{Fig_DFT}
\end{figure}

Here, $R_{\rm B} = {\rm exp}(-\lambda T_{\rm D} m^*/ B)$ is the field induced damping factor where $\lambda = (2 \pi^2 k_{\rm B}/(e \hbar))$ is a constant, $F$ is the frequency of oscillation, and $T_{\rm D}$ is the Dingle temperature. $R_{T}$ is the thermal damping factor given by the LK formula $R_{T} = (X/{\rm sinh(X)})$ where $X = (\lambda T m^* / B)$  and $\beta$ is the phase shift that includes Berry phase information.  The temperature variation of  the FFT amplitude is fitted to thermal damping factor $R_{T}$ which resulted in an effective carrier mass $m^*\sim0.4 m_e$.   From the slope of $ln~[D]$ vs $1/B$ plot, where $D = \Delta R_{xx} B$~sinh(X), we obtain the Dingle temperature ($T_{\rm D}$) as $8$~K (Fig.~\ref{Fig4} (d)).  The carrier life time $\tau =\frac{\hbar}{2 \pi k_B T_{\rm D}}$ and mobility $\mu_{SdH}= \frac{e \tau}{m^*}$ are also estimated. All the parameters obtained from SdH oscillations are listed in Table \ref{Table1}.  To confirm the quasi-$2D$ nature of CaCuSb, we have performed the angular dependence of SdH oscillation, as shown in Fig.~\ref{Fig4}(e), where the angle $\theta$ represents the angle between  magnetic field and current.  However, as given in Ref.~\cite{PhysRevB.92.035123,PhysRevB.99.165133,Zhao2019}, there is another approach of rotation of sample which can also give the Fermi surface information. The angular dependence of the FFT frequency obtained from the experimentally observed quantum oscillations is shown as black filled square in Fig.~\ref{Fig4}(f).  For angles below  $60^{\circ}$, the quantum oscillations are not discernible. This suggests a possibility of an elongated ellipsoidal Fermi surface at the center of the Brillouin zone.  A much larger magnetic field and lower temperature are necessary to observe the SdH oscillations at low angles and oscillations from other Fermi pockets as well. The calculated quantum oscillations using first-principles Fermi surface are shown in Fig.~\ref{Fig4}(f) (see Sec.~\ref{BS_sec} for discussion on Fermi surface). From the detailed band structure calculations, we find that the Brillouin zone contains two hole pockets centered at $\Gamma$ point and small electron pockets located at the $M$ point (see Fig.~\ref{Fig_DFT}). The calculated quantum oscillations frequency for $B~\parallel~[0001]$ is 305~T (40~T) for outer (inner) hole pocket and 160~T for the electron pockets. The extremal area for the outer (inner) hole pocket is $2.9\times10^{18}$~m$^{-2}$ ($3.8\times10^{17}$~m$^{-2}$) and for the electron pocket is $1.5\times10^{18}$~m$^{-2}$. The experimentally observed frequency of 314 T corresponds to an area of $3\times10^{18}$~m$^{-2}$. These results indicate the central large hole pocket at $\Gamma$ point is compatible with the experimental SdH oscillation for $J~\parallel~$[1\={2}10] and $B~\parallel~[0001]$. Furthermore, a detailed comparison of our experimental results shows that the observed Fermi pocket may have an elongated ellipsoidal shape. The anisotropy in the Fermi surface is very well reflected in the large anisotropy in the electrical resistivity (Fig.~\ref{Fig2}(a)) where $\rho_{zz}$ is larger than $\rho_{xx}$, suggesting a quasi-$2D$ character. 

 Background subtracted $\Delta R_{xy}$ (right axis) and $\Delta R_{xx}$ (left axis) are shown in Fig~\ref{Fig5}(a).  It is evident from the figure that quantum oscillations are observed even for the Hall resistance $R_{xy}$, however there is a phase shift of about $90^{\circ}$ between $\Delta R_{xx}$ and $\Delta R_{xy}$.  Interestingly, at sufficiently high magnetic fields ($10$~T~$\leq~B\leq45$~T), the Hall resistivity shows step-like features (plateaus) in the $1/\rho_{xy}$ vs $1/B$ and the plateau size corresponding to two consecutive steps is $\Delta (1/\rho_{xy})/(e^2/h)$ = $7.09 \times 10^{9}$~m$^{-1}$ (Fig~\ref{Fig5}(b)).  In Fig~\ref{Fig5}(b), $n$ refers to the LL index which is obtained from $F/B$ where $F$ is the frequency of the quantum oscillation and $B$ is the field at which $\Delta R_{xx}$ shows minimum in the oscillations. The oscillations observed in Hall data up to a magnetic field of 45~T also have the FFT frequency of 314~T. 
For multilayer quantum Hall effect, the step size of inverse Hall resistivity can be written as $\Delta(1/\rho_{xy} )=Ze^2/h$, where $Z$ is the average $2D$ conduction channels per unit thickness contributing to this multilayer transport.  If we assume these $2D$ parallel conduction channels result in the observed plateau-like feature in the Hall measurement, then the step size in the scale of $e^2/h$ is $\Delta(1/\rho_{xy} )/(e^2/h)=Z\sim 7.09\times10^9$~m$^{-1}$. From the $Z$ value, we can calculate the bulk carrier density $((eF)/h)Z = 5.43\times10^{26}~m^{-3}$ which matches well with the bulk carrier density estimated from the slope of the Hall resistivity data ($5.81\times10^{26}~m^{-3}$). From the HLN model, the obtained average number of conduction channels per unit thickness is $3\times10^{10}$~m$^{-1}$, which is nearly similar to the average number of conduction channels ($Z$) obtained from the Hall plateau. 
Even at such high magnetic fields, no additional frequency or any kind of peak splitting is observed (see Supplementary Materials Fig. S5~\cite{supplement}).  The step-like features observed in Fig.~\ref{Fig5}(b) are something similar to quantized Hall plateaus with high filling factor~\cite{PhysRevLett.118.067702,sasaki1999possibility}. Additionally, the $90^{\circ}$ phase difference between $\Delta R_{xx}$ and $\Delta R_{xy}$, and $\omega \tau = \frac{e B}{m^*} \tau~\sim 11$ for $B = 40~\rm{T}$, indicating the quantum Hall regime. All these suggest that the observed steps in the Hall data may be due to quantum Hall effect.  The linearity in the ($1/\rho_{xy}$) $vs.$~$1/B$ plot (Fig.~\ref{Fig5}(b)) gives additional testimony for the quantization nature of the Hall effect. 
Notably, the quantized Hall steps in $3D$ systems are usually attributed to the $2D$ parallel conduction channels and have been observed in various layered bulk systems such as Bi$_2$Se$_3$~\cite{PhysRevLett.108.216803,Busch2018}, HfTe$_5$~\cite{Galeski2020}, and EuMnBi$_2$~\cite{Masudae1501117}. CaCuSb possesses a layered crystal structure where the Cu-Sb layers are separated by Ca layers. Our experimental results and band structure calculations also indicate that the Cu-Sb layers act as $2D$ conduction channels, which leads to the observation of Hall plateaus in CaCuSb.

\subsection{Band structure of CaCuSb}\label{BS_sec}

First-principles calculations were performed based on the density-functional theory~\cite{kohan_dft} with the projector augmented wave (PAW) method~\cite{paw94}, as implemented in the Vienna $ab initio$ simulation package (VASP)~\cite{kresse96,kresse99}. The generalized gradient approximation (GGA) was used for the exchange-correlation effects~\cite{gga96}. The SOC was considered self-consistently. A plane wave cutoff energy of 420 eV was used and a $10\times 10\times 8$ $\Gamma$-centered $k$ mesh was employed to sample bulk Brillouin zone. The FermiSurfer program was used to visualize the Fermi surface~\cite{Fermisurfer}. We present the calculated band structure CaCuSb in Fig. \ref{Fig_DFT}(a). It is seen to be semimetallic in which small electron pockets located at $M$ overlap in energy with the large hole bands centered at the $\Gamma$ point. The orbital resolved band structure shows that hole bands are dominated by Sb $p$ states whereas the electron bands near the Fermi level are mainly composed of Cu $s$ states. Notably, the bands do not constitute a band inversion, showing that CaCuSb is a nontopological semimetal. The density of states (DOS) in Fig. \ref{Fig_DFT}(b) further reveal a small DOS at the Fermi level which is a reminiscent feature of the semimetallic state. Nevertheless, we find that Ca has vanishing DOS whereas the Cu-Sb has notable DOS at the Fermi level. The Fermi surface is shown in Fig. \ref{Fig_DFT}(c) where the color scale marks the Fermi velocity on each band. It consists of a big ellipsoidal shaped hole pocket with a major axis along the hexagonal $c$ axis and three small electron pockets at the $M$ points. Importantly, the Fermi velocity is higher for the hole bands whereas it is minimum for the electron pockets. This is in accord with the dominant hole-carriers observed in our experimental samples. Furthermore, the finite Cu-Sb states constituted hole bands with relatively higher DOS at the Fermi level. It should be noted that both the Cu and Sb atoms have large SOC which splits the Sb $p_{xy}$ states, leading to two-hole pockets at the $\Gamma$-point. Thus, while the system is topologically trivial, due to inherent large SOC, it can show WAL as seen in our experiments as well as reported widely in other materials~\cite{Hu_2014, BERGMANN19841, Lin_2002}.

\section{Summary}

In summary, the flux-grown CaCuSb single crystal exhibits a huge anisotropy in electrical transport. This anisotropy can be related to the layered crystal structure where the two Cu$-$Sb layers are separated by Ca-layers of the hexagonal unit cell. WAL and quantum oscillations have been observed in the linear and Hall resistivities. The WAL behavior is confirmed from the magnetotransport and low field magnetoconductance analysis. Interestingly, the SdH oscillation is observed in the field range $10 - 14$~T with a frequency at $314$~T, which is primarily due to the large hole-pocket centered at $\Gamma$- point. Low field normalized magnetoconductance falls onto a single curve for different angles and SdH oscillation vanishes below 60$^{\circ}$ suggest the quasi-$2D$ nature of the electrical transport. The high magnetic field Hall resistivity data exhibits step-like features at regular intervals. The angular-dependent magnetoconductance, anisotropy in magnetoresistance, and band structure indicate the presence of multiple $2D$ conduction channels. First-principles calculations have revealed that CaCuSb is a normal semimetal with dominant hole carriers and higher Cu$-$Sb DOS at the Fermi level. A comparison of calculated quantum oscillations with the observed oscillations suggests that the Fermi pockets may have an elongated ellipsoidal shape. It is clear from these results that although CaCuSb possesses a $3D$ layered structure, it can be considered as a pseudo-$2D$ system. The present study would pave way for identifying layered materials as a new platform to investigate quasi-$2D$ transport properties.

\section{ACKNOWLEDGEMENTS} 
We thank Department of Atomic Energy (DAE) of Government of India for financial support. We also acknowledge the support of the LNCMI-CNRS,member of the European Magnetic Field Laboratory (EMFL) for high magnetic field measurements.


%

\end{document}